# Effective reduction in thermal conductivity by high-density dislocations in $SrTiO_3$

Jinxue Ding[1*a], Jiawen Zhang[2a], Jinfeng Dong[3], Kimitaka Higuchi[4], Atsutomo Nakamura[5], Wenjun Lu[2*], Bo Sun[6*], Xufei Fang[1]

[1]Institute for Applied Materials, Karlsruhe Institute of Technology, Karlsruhe 76131, Germany
[2]Department of Mechanical and Energy Engineering, Southern University of Science and Technology, Shenzhen 518055, China
[3]School of Materials Science and Engineering, Nanyang Technological University, Singapore 639798, Singapore
[4]Institute of Materials and Systems for Sustainability, Nagoya University, Nagoya, Aichi 464-8601, Japan
[5]Graduate School of Engineering Science, Osaka University, Osaka, Toyonaka 560−8531, Japan
[6]Tsinghua SIGS and Guangdong Provincial Key Laboratory of Thermal Management Engineering & Materials, Tsinghua University, Shenzhen 518055, China

[a]These authors contribute equally to this work.

*Corresponding authors:
jinxue.ding@kit.edu (J. D.), luwj@sustech.edu.cn (W. L.), sun.bo@sz.tsinghua.edu.cn (B. S.)

**Abstract:**

Decreasing thermal conductivity is important for designing efficient thermoelectric devices. Traditional engineering strategies have focused on point defects and interface design. Recently, dislocations as line defects have emerged as an additional tool for regulating thermal conductivity. In ceramics-based thermoelectric materials, the key challenge lies in achieving sufficiently high-density dislocations to effectively scatter phonons, as the typical dislocation density in ceramics after bulk deformation is constrained to ~$10^{12}$ $m^{-2}$. In this work, we adopted the mechanical imprinting method and achieved a dislocation density of ~$10^{15}$ $m^{-2}$ in single-crystal $SrTiO_3$, which is known for its room-temperature plasticity and acts as a promising material for thermoelectric applications. Using the time-domain thermoreflectance (TDTR) method, we measured a ~50% reduction in the thermal conductivity over a broad temperature range (80−400 K) with the engineered high-density dislocations. These results suggest that tuning dislocations could offer an alternative path to minimizing thermal conductivity for engineering thermoelectric materials.

Regulating thermal transport is of critical importance for a wide range of applications, including thermal management, energy conversion, and functional material design[1–3]. In contrast to electrons, phonons, the dominant heat carriers in crystals, are electrically neutral and lack mass, rendering them unresponsive to external fields such as electric or magnetic stimuli. Consequently, modulation of phonon transport is predominantly achieved through structural modifications across multiple length scales, a strategy known as defect engineering. These include zero-dimensional (0D) point defects such as vacancies or interstitial atoms, 1D dislocations, 2D planar defects such as boundaries and stacking faults, and 3D precipitates or secondary phases. Among these approaches, dislocation engineering has attracted growing interest, driven by recent advances in mechanically inducing high-density dislocations (up to $10^{14-15}$ $m^{-2}$) in functional ceramics, particularly perovskite oxides[4,5]. Theoretical modeling results in $SrTiO_3$ (STO) by Johanning et al.[6] suggest that such high dislocation densities can effectively reduce thermal conductivity and, in some cases, even improve electrical properties by dislocation-induced space charge fields[7,8], making them particularly advantageous for thermoelectric (TE) applications[9,10].

In recent years, the role of dislocations in enhancing phonon scattering in ceramics has been revisited with promising proofs of concept. For example, Yin et al.[11] introduced high-density dislocations (up to ~$9.1\times10^{16}$ $m^{-2}$) into BiCuSeO by ultrahigh-pressure sintering to strongly suppress the phonon transport. Abdellaoui et al.[12] studied the characteristics of dislocations in PbTe-based TE material, including their distribution, alignment, and local chemistry, using correlative microscopy techniques. Beyond TE materials, the phenomenon of dislocation-tuned thermal conductivity has also been observed and investigated in semiconductors and other functional ceramics. The anisotropic thermal conductivity caused by oriented dislocations was first predicted by Klemens[13] in the 1950s and more recently experimentally verified by Sun et al.[14] in InN films with oriented threading dislocations. Khafizov et al.[15] highlighted that dislocation loops in nuclear ceramics ($CeO_2$), formed due to irradiation damage, can significantly affect heat transport.

Thermal conductivity can depend critically on dislocation density.[16] At low dislocation densities, where the spacing between dislocations is much larger than the phonon mean free path (MFP), their influence on phonon transport becomes negligible. In 2020, Johanning et al.[6] engineered dislocations into bulk STO by high-temperature plastic deformation. However, no enhanced phonon scattering was observed in the plastically deformed samples due to the low dislocation density (~$10^{12}$ $m^{-2}$). At that time, achieving a high dislocation density of around $10^{14-15}$ $m^{-2}$ in STO was not feasible. Recent advancements have started to overcome these challenges, making it possible to engineer high-density dislocations in ceramic oxides. However, despite this progress, direct experimental evidence showing that mechanically induced dislocations reduce thermal conductivity remains exclusive.

To address this open question, here we adopt undoped single-crystal $SrTiO_3$ (to avoid the complexity of dopants and grain boundaries), which exhibits exceptional large plasticity even at room temperature[17] and is also a well-established material in the TE field[18,19], making it an ideal candidate for this research. With the mechanical deformation toolbox developed by the present authors[4,20,21], we engineered a high dislocation density (up to ~$10^{15}$ $m^{-2}$) in the sample by near-surface mechanical deformation. This approach allows to evaluate the thermal conductivity in the dislocation-rich region via the time-domain thermoreflectance (TDTR) method.[22–24]

Undoped STO single crystals with a geometry of 5 × 5 × 1 $mm^3$ from Alineason Materials Technology GmbH (Frankfurt am Main, Germany) were used. Dislocations were engineered in the near-surface region using a mechanically seeded dislocation approach through grinding and polishing (schematic in **Fig. 1a**). Grinding was performed using SiC papers (P2500), applying a force of 15 N and a rotation speed of 150 rpm. The following polishing steps were carried out to ensure a smooth surface. More details on this dislocation seeding approach can be found elsewhere.[4] In what follows, pristine STO is referred to as *p*-STO, and STO with surface dislocations as *d*-STO. To observe the dislocation substructure near the surface using the high-voltage transmission electron microscopy (HVTEM), the conventional TEM foil preparation method, which does not introduce additional dislocations into the sample, was adopted. First, two samples with well-preserved surfaces after polishing were prepared and bonded with epoxy to form a stable interface. Then, while ensuring the integrity of the interface, normal mechanical polishing, dimpling, and ion milling with Ar ions were performed from the side to produce a thin film for TEM observation that retained the original surface structure. The observation images were obtained using an ultra-high voltage electron microscope (JEOL JEM-1000K RS, 1000 keV) in bright-field mode.

For local dislocation structure characterization and 3D reconstruction of the dislocations, additional TEM thin lamellae with approximately 50 nm in thickness were prepared using a focused ion beam (FIB) system (Helios Nanolab 600i, FEI, Hillsboro, USA). These TEM samples were characterized using a field emission TEM (Talos F200X G2) operating at 200 kV. For annular bright-field (ABF)-scanning TEM (STEM) imaging a probe semi-convergence angle of 17 mrad and inner and outer semi-collection angles ranging from 13 to 21 mrad were utilized. The microstructures of STO, including dislocations, were analyzed using a 200 kV TEM in STEM mode. Tilt-series of ABF-STEM images were acquired using Xplore3D software (FEI Company), with dynamic focusing incorporated during image acquisition. A high-angle tomography holder (Model 2020, Fischione, USA) was used for tilt-series acquisition, with a maximum tilt angle of 60°, which was necessary to observe the same set of dislocations in all tilt-series images. During these experiments, the [001] direction of the lamella specimens was aligned with the tilt axis of the specimen holder. The tilt-series images were then aligned, and 3D datasets were reconstructed using Inspect3DIII software (Thermo Fisher Scientific Inc.), employing the simultaneous iterative reconstruction technique (SIRT) for 40 iterations. All tilt-series STEM images were utilized for the 3D reconstruction, and the final 3D models were visualized using Amira-Avizo™ software.

To measure the through-plane thermal conductivity in the near-surface region, an aluminum (Al) layer with a thickness of ~70 nm was deposited on the surface of samples by e-beam evaporation. The TDTR method was carried out using the 5x objective lens (1/e$^2$ radius of 12 µm) with a modulation frequency ($f$) of 10 MHz. Thermal transport is modeled as the propagation of thermal waves in layered structure at a given modulation frequency, $f = 10$ MHz in this case. This results in a thermal penetration depth $d$, which is given by the equation: $d = \sqrt{\Lambda/\pi f C}$, with $\Lambda$ and $C$ being the thermal conductivity and volumetric specific heat of the sample, respectively. The penetration depth $d$ is calculated to be less than 1 µm, to make sure we measure dislocation-rich region.

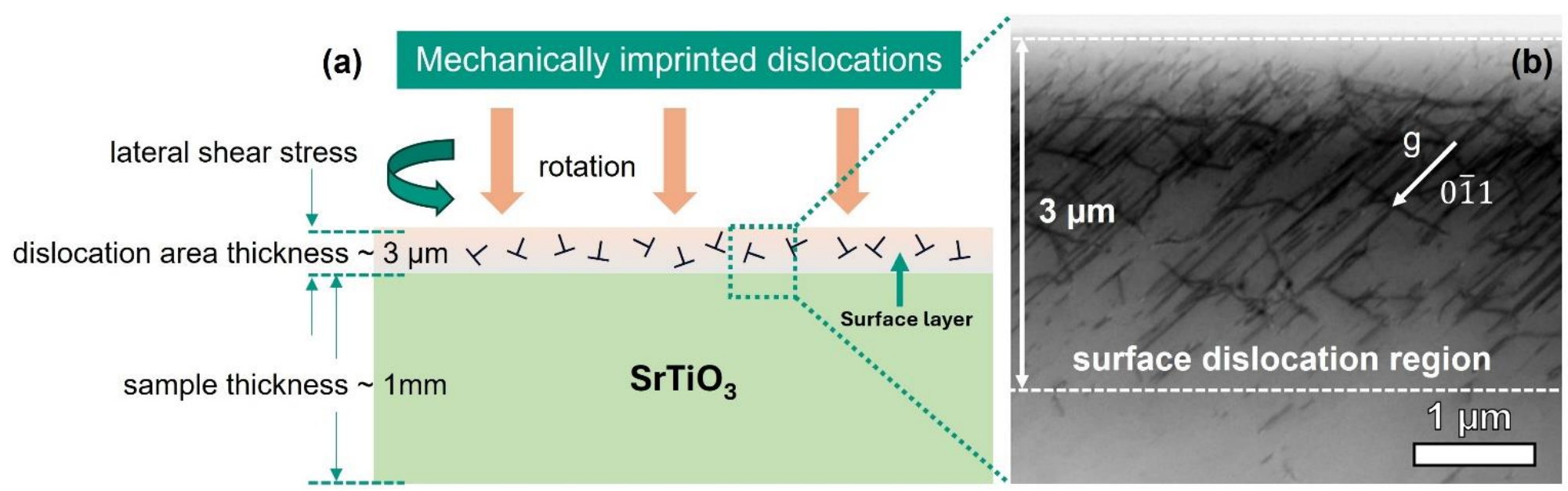

**FIG. 1.** (a) Schematic illustration of the mechanical imprinting method; (b) cross-sectional HVTEM image of near-surface region of the *d*-STO sample, as indicated by green rectangle in (a). The upper dashed line in (b) indicates the sample surface.

The HVTEM image in **Fig. 1b** revealed that the cross-section of the deformed surface layer (up to a depth of ~3 µm) is enriched with high-density dislocations. In contrast, bright-field TEM (BF-TEM) image for *p*-STO shows no visible dislocations (**Fig. S1**). Particularly, no subgrain boundaries or amorphous phases were detected in the highly deformed region in *d*-STO, according to the selected area electron diffraction (SAED) pattern (**Fig. S1**). The dislocation density is determined by the line intercept method using TEM images[25] (**Fig. S1c** and **Fig. 1b**), reaching up to ~$10^{15}$ m$^{-2}$. As the grinding particles have a limited range of stress field exerted on the sample surface, the dislocation density drops sharply beyond ~3 µm in depth. Beyond this deformation layer (bottom of **Fig. 1b**), the sample can be regarded as pristine STO.

The pristine sample has a grown-in dislocation density of $10^{10}$ m$^{-2}$, as reported in pervious studies[26,27]. The limited dislocation region in the depth direction merits the use of TDTR measurement, which can achieve a penetration depth less than 1 µm to focus on the impact of dislocations.

The temperature-dependent thermal conductivity demonstrates a clear difference (**Fig. 2**) for both *p*-STO and *d*-STO samples. For benchmarking purpose, in *p*-STO with a dislocation density of ~$10^{10}$ m$^{-2}$ (almost dislocation-free for the laser spot area), the TDTR results are in excellent agreement with those reported by Yu et al.[28], measured using a 3$\omega$ method. The *p*-STO sample exhibits a typical temperature-dependent behavior of thermal conductivity. As the temperature increases within the measurement range, thermal conductivity initially rises due to the enhanced phonon population and the corresponding increase in heat-carrying phonons. As the temperature continues to rise, phonon-phonon scattering intensifies, leading to a peak in thermal conductivity, after which it decreases. For *d*-STO, ~50% to ~60% decrease in thermal conductivity is observed over the wide temperature range (80−400K). The distinct difference between the *p*-STO and *d*-STO samples underscores the significant role of dislocations in scattering phonons, effectively suppressing temperature-dependent behaviors and reducing thermal conductivity. As presented in **Fig. S2**, the *p*-STO sample displays a typical TDTR signal with one picoacoustic peak, indicating that the thermal stress wave is reflected at the interface between the Al film and STO. In contrast, the *d*-STO sample shows a second peak, suggesting the presence of acoustic wave reflections caused by dislocations.

For both *p*-STO and *d*-STO samples in the measured temperature range, heat transport is predominantly governed by phonon transport and the contribution of electronic thermal conductivity is negligible (see Supplementary Materials). Therefore, measured thermal conductivity can be considered as $\Lambda_l$. The theoretical minimum of $\Lambda_l$ for highly disordered crystals can be calculated using the Cahill-Pohl model[29], which is expressed as follows:

$$\Lambda_{l,min} = \left(\frac{\pi}{6}\right)^{\frac{1}{3}} k_B n^{\frac{2}{3}} \sum_i v_i \left(\frac{T}{\theta_i}\right)^2 \int_0^{\frac{\theta_i}{T}} \frac{x^3 e^x}{(e^x - 1)^2} dx \quad (1)$$

where $x = \frac{\hbar\omega}{k_B T}$ with $\omega$ being the phonon angular frequency, $k_B$ is the Boltzmann constant, $v_i$ the sound velocity, $n$ is the number density of atoms, and $\theta_i$ is the Debye temperature. The calculated $\Lambda_{l,min}$ ranges from 1 to 2 Wm$^{-1}$K$^{-1}$, as shown in **Fig. S3**. It suggests that the introduced high-density dislocations are not sufficiently disordered to achieve the lower limit of $\Lambda_l$, which is reasonable that the theoretical $\Lambda_{l,min}$ is achieved with a phonon MFP down to the distance of the nearest atoms.

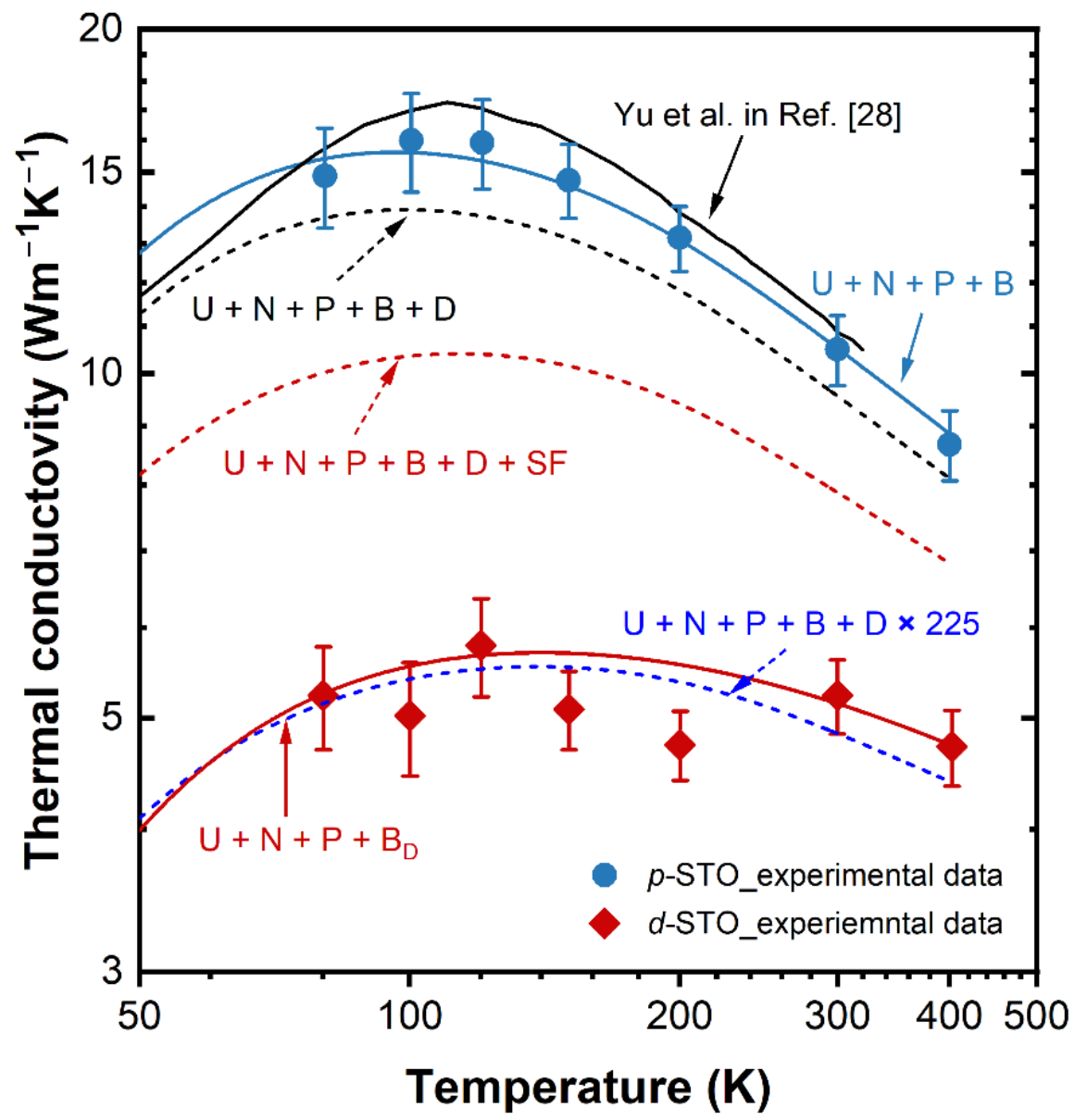

**FIG. 2.** The measured temperature-dependent thermal conductivity by TDTR and analysis of phonon transport process using modified Debye-Callaway model. Phonon-dislocation scattering is evaluated by treating dislocations as line defects (D) and/or 2D stacking faults (SF), following the Klemens model. Note: U for Umklapp phonon-phonon scattering, N for normal phonon-phonon scattering, P for point defect scattering, B for boundary scattering, D for dislocation scattering, SF for stacking faults scattering, $B_D$ for dislocation-induced boundary scattering.

The experimental data indicate that dislocations significantly suppress $\Lambda_l$, especially at low temperatures. To gain deeper insights into the role of dislocations, we employ the modified Debye-Callaway model to interpret the measured data. Then, $\Lambda_l$ can be calculated by considering the contribution from individual acoustic phonon branches (one longitudinal and two transverse acoustic branches), as expressed by the following equations[30,31]:

$$\Lambda_l = \sum_i \Lambda_{l,i} = \sum_i \frac{k_B}{6\pi^2 v_i}\left(\frac{k_B T}{\hbar}\right)^3 \left(I_{i1} + \frac{{I_{i2}}^2}{I_{i3}}\right) \qquad (2)$$

The details for a specific phonon branch ($I_{i1}$, $I_{i2}$, $I_{i3}$) are presented in the Supplementary Materials. The combined relaxation time for phonons ($\tau_i^C$) can be written as follows according to Matthiessen's rule, including normal phonon-phonon scattering (N), Umklapp phonon-phonon scattering (U), point defect scattering (P), boundary scattering (B), dislocation scattering (D), etc., as in:

$$\frac{1}{\tau_i^C} = \frac{1}{\tau_i^N} + \frac{1}{\tau_i^U} + \frac{1}{\tau_i^P} + \frac{1}{\tau_i^B} + \frac{1}{\tau_i^D} + \cdots \tag{3}$$

It may seem odd at first sight for including point defect scattering and boundary scattering for single-crystal *p*-STO. According to the model (see Supplementary Materials), these two mechanisms should be considered even for the undoped single-crystal sample, as phonon scattering by defects typically dominates thermal transport below the Debye temperature. When the contribution of boundary scattering is ignored, $\Lambda_l$ can be very high at low temperatures, as shown in **Fig. S3**. Considering that STO goes through a phase transition from cubic to tetragonal around 105 K[32], the results suggest that domain structures[33,34] occurred near the phase transition may behave like 2D boundaries and can strongly scatter phonons. The point defect scattering parameter $\Gamma$ and boundary scattering characteristic dimension $L_c$ are fitted based on the experimental data of *p*-STO, yielding values of $\Gamma$ = 0.0206 and $L_c$ = 458 nm, which are then applied to *d*-STO. The point defect scattering can be explained by the well-known existence of oxygen vacancies[35], locally atomic-scale structures caused by off-centering atoms[36,37], and other impurity ions introduced during the growth of single-crystal STO.

However, the Klemens model[13] for dislocation scattering predicts only a small reduction in thermal conductivity, as indicated by the black dashed line in **Fig. 2**. It should be noted that this model does not account for the potential reduction in phonon velocity caused by stress-induced lattice softening, which may explain part of the discrepancy but not the large deviation observed. This large discrepancy may arise since these models did not consider the complex dislocation structures, as visualized by the ABF-STEM observations (**Fig. 3**). The classic models assume dislocations as weak and isolated phonon scattering centers, more valid for low-density dislocations ($10^{12-13}$ $m^{-2}$). With dislocation density increasing up to ~$10^{15}$ $m^{-2}$, they are no longer isolated line defects but form a complex spatial network, as depicted in **Fig. 3f** for the reconstructed 3D tomography. Five sAnapshots from the ABF-STEM video (Multimedia View **Video S1**), captured while rotating the sample, are displayed in **Figs. 3a**-**3e**, respectively. These images reveal that, in addition to typical 1D dislocations lines, other defects, e.g., dislocation loops, stacking faults, dislocation dipoles, all arranged in a sophisticated 3D manner, are also present. These complexity of dislocation structures and their interaction (strain field) with each other shall be considered in the future model to quantitively describe the dislocation-induced thermal transport behavior.[38]

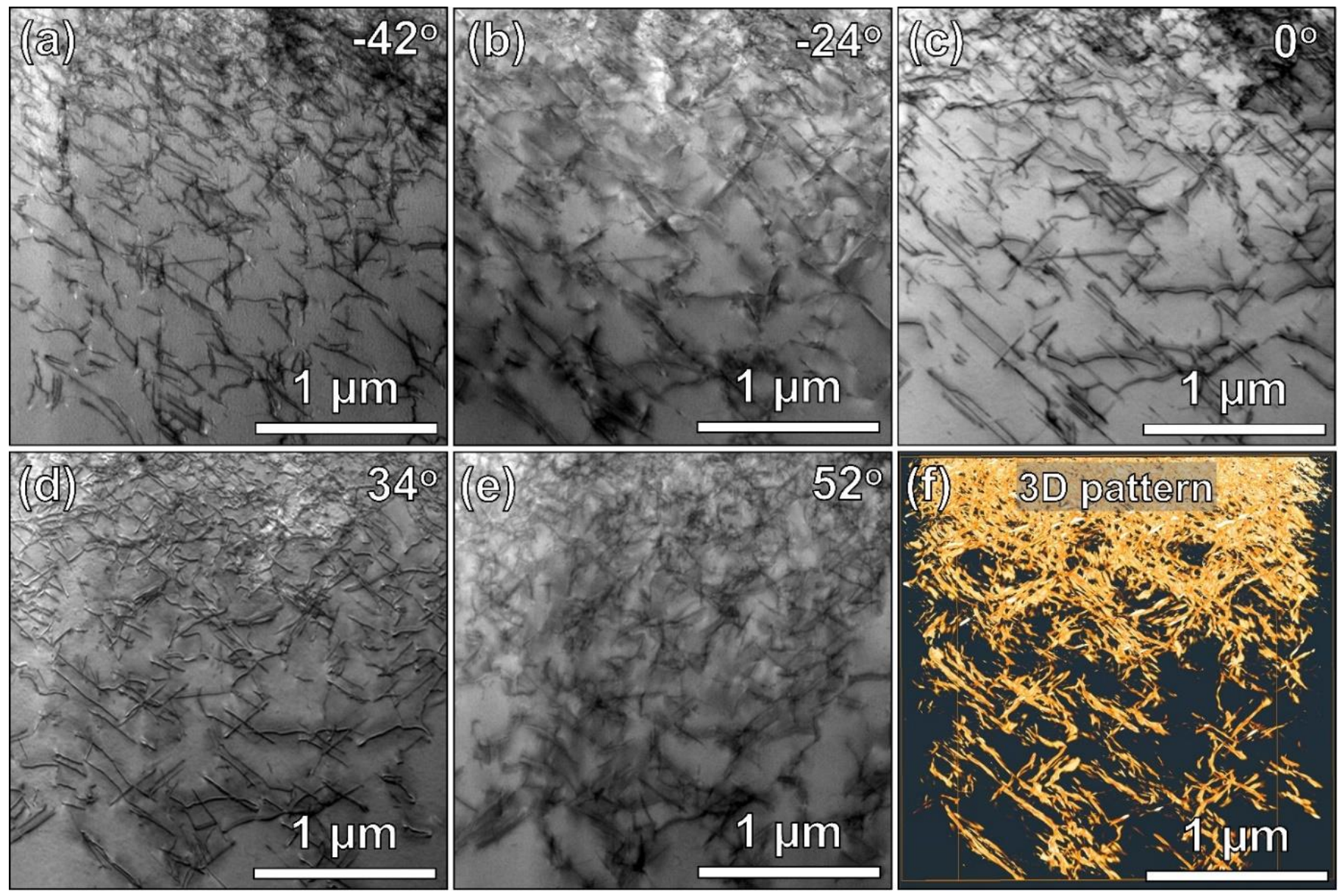

**FIG. 3.** 3D reconstruction (Multimedia View) of *d*-STO lamella with mechanically seeded dislocations. (a-e) A series of ABF-STEM images tilted along <100> axis from -42° to 52°; (f) the corresponding 3D image showing the dislocation networks.

Note that back in 1959, significant discrepancies were observed between the experimental data and values predicted by Klemens model in LiF.[39] Carruthers[40] later found that the Klemens model underestimated the influence of strain fields from edge-type dislocations and proposed the later known Carruthers model. Instead, for screw-type dislocations, Carruthers model concurred with the predictions made by Klemens. We note that the Carruthers model may not apply to the experimental data here, as the surface dislocations mechanically introduced are primarily screw-type dislocations, with very few edge-type dislocation segments located at the end of the screw dislocation lines[4]. When stacking fault (SF) scattering is included, the predicted thermal conductivity decreases further (red dashed line in **Fig. 2**), although it may still not reach the levels observed in the measured data. To fit the experimental data, the dislocation scattering rate from Klemens model needs to be magnified by a factor of 225, indicated by the blue dashed line in **Fig. 2**. Although the physical justification for the prefactor of 225 remains unclear, it is just used as a fitting parameter to match the measured data. The contributions of each scattering mechanism and the spectral lattice thermal conductivity (see Supplementary Materials **Fig. S6**) demonstrate that dislocations, as crystal defects, primarily scatter low-frequency phonons, with minimal impact on high-frequency phonons.

Given the complexity of dislocations in the high-density regime, a cell-like dislocation network may form a quasi-periodic boundary. By treating dislocation scattering as dislocation-induced boundary scattering ($B_D$), the experimental data can be well captured by adjusting the characteristic dimension parameter $L_c$ to be 100 nm (larger than the average dislocation spacing ~30 nm), as shown by the red solid line in **Fig. 2**. This characteristic length $L_c$ should be closely related to the density and arrangement of dislocations, and potentially other physical properties (e.g., elasticity, crystal structure, etc.) of the material, which can affect dislocation interactions. It could be valuable to link $L_c$ to other parameters in the future work, as this may provide insights into the impact of dislocations on phonon scattering, particularly in the high-density regimes. Moreover, for high-density dislocations, caution must be taken for the determination of their density and spatial arrangement since dislocations may overlap or form entangled networks (**Fig. 3**). Furthermore, the high-density dislocations in STO may interact with domain structures at temperatures below 105 K, leading to the formation of more complex structures. Thus, more advanced correlative techniques to comprehensively determine dislocation structures are needed for future studies. For example, dark-field X-ray microscopy can provide both high spatial resolution and the ability to directly probe the dislocation structures in 3D manner, yet its spatial resolution is limited to ~100 nm.[41]

In summary, mechanically seeded high-density dislocations, up to ~$10^{15}$ $m^{-2}$ as visualized by transmission electron microcopy, in undoped single-crystal $SrTiO_3$ is experimentally observed to significantly reduce thermal conductivity by up to ~60%. The impact of dislocations on reducing thermal conductivity is much more pronounced than predicted by previous models, particularly at low temperatures where low-frequency phonons dominate thermal transport, resulting in a reduced temperature dependence of thermal conductivity. The large discrepancy is likely caused by the complex dislocation structures and spatial distribution, which were not accounted for in the past models. The interactions between dislocations and phonons require further exploration to gain a deeper understanding of the role of dislocations in tuning thermal conductivity. The influence of dislocations on electrical transport properties is also considerable and deserves further investigation, particularly for thermoelectric applications. The high-density dislocation imprinting method in functional ceramics provides an opportunity to deepen our understanding of dislocation-related phenomena.

**Supplementary Material**

See the supplementary material for detailed information on the surface dislocation structure, dislocation density measurements, TDTR experiments, lattice thermal conductivity calculations, the influence of boundary scattering on lattice thermal conductivity, and the analysis of dislocation-phonon interactions.

**Acknowledgement**

J. Ding and X. Fang thank for the final support by ERC project MECERDIS (No. 101076167). Views and opinions expressed are however those of the authors only and do not necessarily reflect those of the European Union or the European Research Council. Neither the European Union nor the granting authority can be held responsible for them. W. Lu acknowledges the support by Shenzhen Science and Technology Program (grant no. JCYJ20230807093416034) and the Open Fund of the Microscopy Science and Technology-Songshan Lake Science City (grant no. 202401204). The authors acknowledge the use of the facilities at the Southern University of Science and Technology Core Research Facility. B. Sun would like to thank the National Science Foundation of China (Nos. 52161145502 and 12004211), the Shenzhen Science and Technology Program (grant no. RCYX20200714114643187 and WDZC20200821100123001), and the Guangdong Special Support Program (No. 2023TQ07A273). We thank Dr. L. Porz from Illutherm GmbH, Prof. G. Snyder from Northwestern University, Prof. J. Rödel from TU Darmstadt, Prof. M. Yoshiya and W. Sekimoto from Osaka University for the initial discussions. The authors also thank S. Stich and H. Song for sample preparation.